# Water-Walled Microfluidics Makes an Ultimate Optical Finesse

Shai Maayani[1], Leopoldo L. Martin[1], and Tal Carmon[1]*

[1]Technion-Israel Institute of Technology, Haifa 32000, Israel

*Correspondence to: tcarmon@technion.ac.il

**Liquids serve microcavity research ever since Ashkin's studies on optical resonances in levitating droplets[1] to recent optofluidic[2-6] resonators[7-15]. Droplets can provide optical quality factor (Q) in proximity to the limit restricted by water absorption and radiation loss[16]. However, water μdrops[7] vaporize quickly due to their large area/volume ratio. Here we fabricate a water-air interface[17] that almost entirely surrounds our device, allowing for >1,000,000 recirculations of light (finesse). We sustain the droplets for >16 hours using a nano-water-bridge that extends from the droplet to a practically-unlimited distant-reservoir that compensates for evaporation. Our device exhibits surface tension 8000-times stronger than gravity that self-stabilizes its shape to a degree sufficient to maintain critical coupling[18] as well as to resolve split modes[19]. Our device has 98% of their surrounding walls made strictly of water-air interfaces with concave, convex or saddle geometries, suggesting an arbitrary-shape microfluidic technology with water-walls almost all-over.**

Water walls were used in various applications[20] including reactions at the gas-liquid interface[17]. Here we curve such walls to form a stable spheroidal shape (Fig 1c) which can benefit optical resonators where light circumferentially circulates in water next to their interface with air. When circumference is an integer number of wavelengths, this circulating light is resonantly enhanced and called a whispering gallery mode.

The current state of the art in optofluidic resonators[3,5,6] include droplets falling near a laser pulse[7], pipettes[12], toroids[9,14,21], as well as submerged spheres[8], droplet lasers[13], water droplets in oil[11] and water-glycerol on a hydrophobic surface[15]. Such resonators can benefit from a variety of techniques, among them are cavity-enhanced spectroscopy[22], cancelling frequency drifts in the resonator[21] and in the pump[14] by using a differential detection scheme, broad-band spectroscopy using frequency combs[23], and recently, exploiting ring-up for ultrafast spectroscopy[24]. In their inherent essence, micro-cavities confine multiple optical passes. Accordingly, and with several close descriptions encountered in literature, the major goodness-of-fit parameters of such resonators are their smallness, $1/V_m$[19], where $V_m$ is the volume of the optical mode, their optical quality factor, Q, and the optical intensity, $f^2(\mathbf{r})$, at the region of interest, $\mathbf{r}$[19].

Water surrounded by air that we are using here, suggests the best of several worlds. This includes (a) a maximal contrast between air and the water-core for tight confinement ($1/V_m$) while suffering a minimal radiation-loss penalty[16]. Additionally, (b) air is so transparent that the water-droplet's Q can be limited by water absorption only. Lastly, (c) optical intensity, $f^2(\mathbf{r})$, is almost fully contained in the water near their interface with air. That said, thermal capillary waves[25] can scatter resonantly enhanced light near the interface[26].

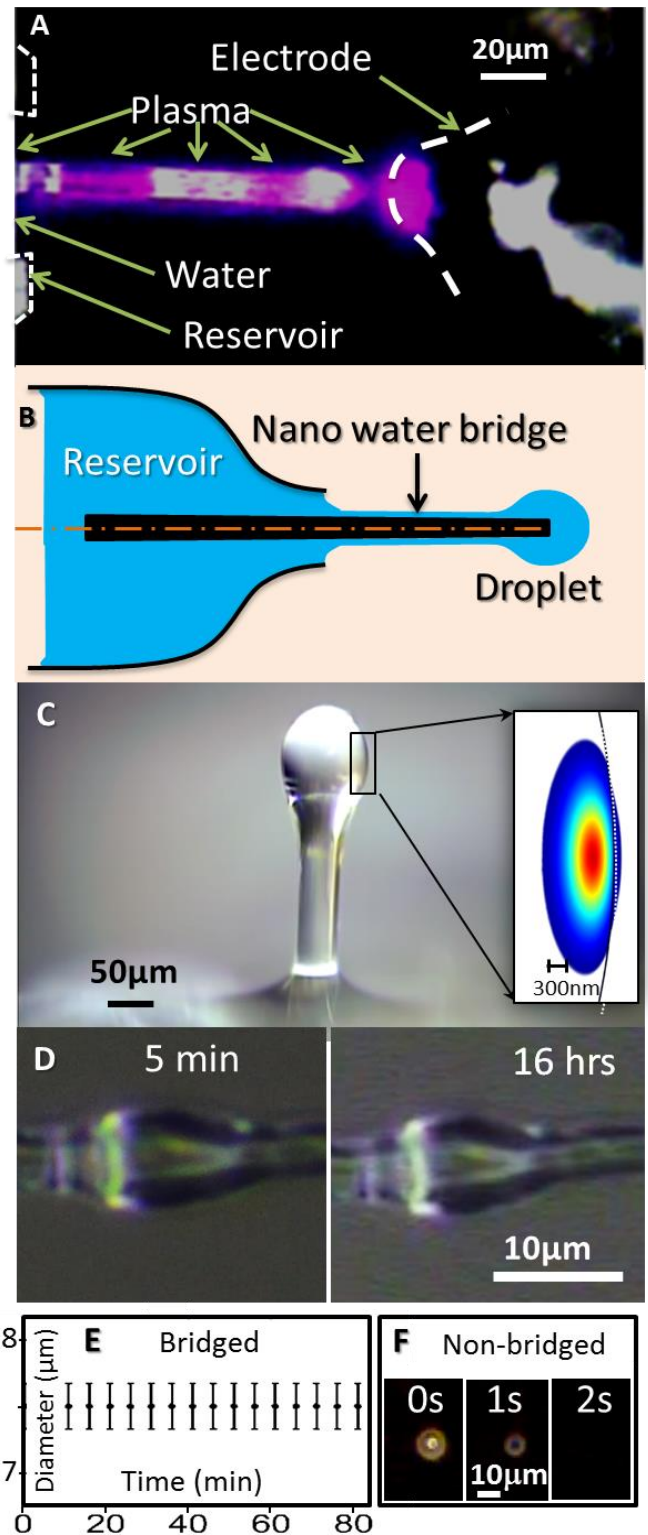

**Fig. 1. Droplet sustainer**. **(A)** A silica stem is **plasma treated** in order to modify its hydrophilicity. **A water microdroplet** is then formed near the stem's end **(B right)** while in fluidic contact with a **water reservoir (B left)** via **a nano water bridge** covering the stem **(B center)**. **(C) Micrograph of the drop** together with its calculated mode (inset). **(D-E) Monitoring the drop size** where error bars represent microscope resolution. **(F) Control group**: Evaporation of a droplet on a plate. Vertical and horizontal configurations were equivalent here, since gravity is negligible when compared to surface tension.

We will now explain our technology for *supporting a durable microfluidic device that is surrounded by water walls from most of its sides*. Fig. 1.b-c describes our droplet sustainer, where a nano water-bridge extends from the droplet all the way to a practically-unlimited reservoir that feeds it. Total compensation for evaporation is achieved by designing the hydrophilicity and geometry of our solid surfaces, to enable this bridge; while bearing in mind that water, from energy consideration, prefers to minimize the product of interfacial-tension and area. A hydrophobic pipette was used to prevent leakage from its small hole, but then water inside could not reach its thin end. To solve this problem, an inner hydrophilic filament is used to enable guidance of water to its narrow end. As will be explained in more details in the method section, a wet-plasma treated[27] (Fig 1.a) cylindrical stem at the thin end of the pipette is holding the microdroplet (Fig 1.b-c right) while in fluidic contact with the water bridge (Fig 1.b-c center). The micrograph of our sustainer presents a droplet (Fig 1.d), where light can circumferentially circulate along an equatorial line to form an optical whispering gallery resonance. Upon need, water can surround the full length of the stem (Fig 1.d) including the formation of a droplet at its end. The existence of the water bridge is confirmed by touching it with a dry object and watching it absorb liquid.

The droplet survives perturbations such as holding it in a hand (from its sustainer) and walking with it to another setup. The droplet also survived touching it with the 1 micron tapered fiber. In fact, the taper broke in such tests while the droplet lived on. This is expected since surface tension governs at small scales. To put it into proportion, surface tension in our device is more than 8000-times stronger than gravity as defined by its Eötvös number (See Supplementary Information), meaning that such droplets might even withstand high accelerations. While regular droplets at these scales quickly vaporize (Fig. 1.g and SI Movie 2), our microfluidic device was sustained for more than 16 hours (Fig 1.e); when we had to interfere and cut the experiment, in order to continue to the next study.

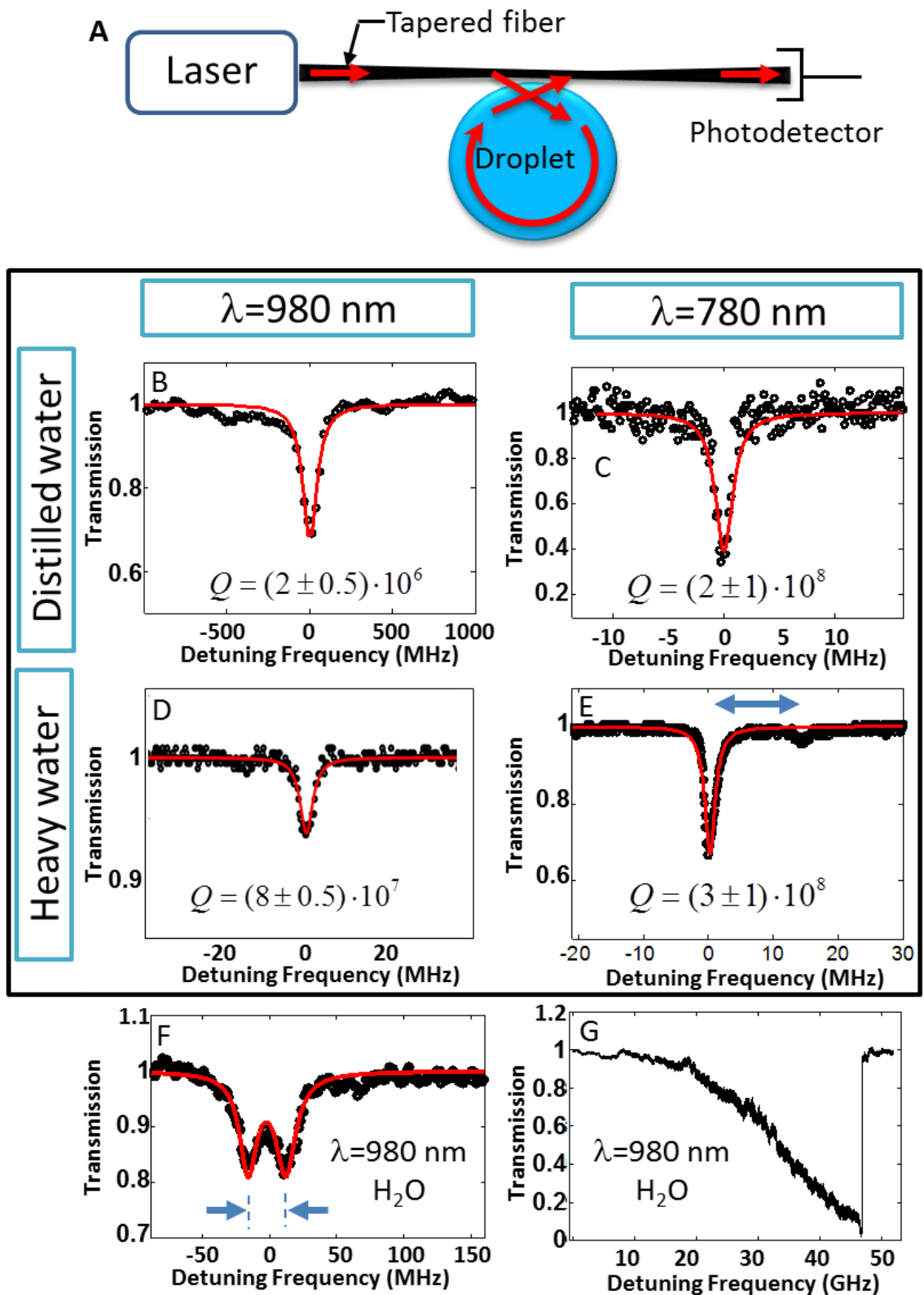

**Figure 2: Optical resonances in micro water droplets.** **(A)** Experimental setup **(B-E)** Monitoring **resonator transmission** while scanning the wavelength through resonance provides the linewidth. **(E, F) Splitting between counter-propagating droplet modes** was commonly resolved as marked by the arrows **(G) Coupling efficiency close to 99%** was measured at critical coupling conditions. The droplet's diameter was between 20 and 40 μm.

We can now proceed to demonstrate our claim of *having high-Q optical resonances in this sustainable water drop*. We used a tapered fiber[11,15,28] to evanescently couple light in and out of the resonator (Fig 2.a). We deduce the optical quality factor of our micro-droplet resonator from the measured optical linewidth ($\Delta\lambda$) by using $Q=\lambda/\Delta\lambda$. As it is important to prevent linewidth narrowing[29] in such measurements, we were reducing the optical power and working at the undercoupled regime. Additionally, our Q measurements were performed at the broaden scan and not at the narrowed one. Heavy and distilled water droplets were tested at red and near infrared to provide Q as shown in figure 2.b-e. Both water and heavy water supports an optical finesse >million at 780 nm. As seen when diving in the ocean, blue light penetrates water much deeper than red. Thus, 480 nm light will enable one, or maybe even two, orders of magnitude improvement in our measured finesse. Still, even at 780 nm, with an optical mode almost completely overlapping with water, our finesse-overlap product is more than 1000 times higher than other devices. This said, finesse higher than ours was demonstrated using solid devices[30]. As for the interaction of light with thermal capillary waves[26], our finesse measurement (Fig 2 C, E) reveals that for a mode as calculated in figure (Fig 1C inset) attenuation via scattering from Brownian capillary waves is smaller than 0.0001 cm$^{-1}$.

Trying to estimate our droplet's stability from its resonance fluctuations, we analyze the resonance shape that appears in Fig 2 E which is 5% distorted when compared to a perfect Lorenzian (red). Knowing that drifts in the resonance wavelength proportional to the deviations in cavity radius, we calculate about 11 nm/s radius fluctuations in our resonator. While further studies of cavity stability are still needed, we can say, at this stage, that stability and linewidth are practically sufficient to resolve splitting between the counter-circulating modes[19] (Fig 2.e-f, arrows). Such splits were shown in the past to facilitate differential detection of the relative drift between the two modes while other drifts are canceled out[21].

Lastly, we establish critical coupling by bringing the taper closer to the droplet. As a result, transmission at resonance was dropping to almost zero (Fig 2.g), indicating that the coupling efficiency to our devise is nearly 99%[18]. Contrary to common solids, the thermal broadening[29] is seen here (Fig 2.g) while scanning towards the shorter wavelengths. This is expected since the thermal coefficient of refractive index for water is negative.

In conclusion, devices bounded by water-air interfaces can host optical resonances on their edges. Until today, the fast evaporation of the entire device was challenging such experiments. This technology stopper is mitigated here by establishing a nano water bridge that feeds the device to enable water-walled almost all-over microfluidics at room pressure and temperature, with no controller, no feedback loop, and no additives to the water.

**Methods:** Vertical and horizontal configurations were equivalent here, since gravity is negligible when compared with surface tension.

**Building the droplet fabrication setup:** A glass pipette (WPI, TW100-4) is tapered to an inner diameter of 200 μm while heated by a hydrogen flame. A silica fiber (Corning, SMF 28) is similarly tapered to form a 10 μm diameter cylinder. We then insert the cylinder inside the pipette as seen in figure 1 a-b, and fill the pipette with water, which acts as a water reservoir. The pipette has an inner hydrophilic filament (Extended Data Fig. 1.d) which makes it easier for water to reach its thin side. At the same time, the pipette body is made of a less hydrophilic material, which makes it difficult for the droplet to spontaneously drip as typical for such shapes[31].

Voltage is applied by using two electrodes, one dipped in water (inside the pipette) and the other (platinum electrode) in air (Extended Data Fig. 1b). The voltage is increased until obtaining break down to plasma between the electrode and water (at about 1000 V/mm). The voltage is turned off about one second after the plasma is generated. Such plasma modifies the contact angle between water and silica as described in[27]. The setup is now ready for providing micro droplets upon need and the process described above, including the plasma treatment, should not be repeated.

**Making a sustainable droplet:** In order to pull a water droplet out of the pipette reservoir, we apply low voltage via the same electrodes used in the previous paragraph. This time, the voltage is kept below what is needed for plasma. The drop comes out of the pipette all the way to the end of the cylinder so that water are fully covering cylinder (Extended Data Fig. 1.c) and our device is bounded by free water walls from all of its direction except for the thin stem holding it. The droplet side that is closer to the pipette is water bridged to the pipette reservoir while having a meniscus shape typical for such boundaries[32]. The voltage is now turned off and the drop is sustained (at zero voltage) as long as long as there is water in the feeding reservoir. To summarize this paragraph, making a sustainable droplet, as explained here, involves only turning on and off a relatively low voltage; which takes about a second.

Failure in fabrication occurred only when we were using a very clean system, probably because water was not properly conducting when clean.

**Supplementary Information**

The Eötvös number is a dimensionless number that is relevant to our system as it compares gravitational forces to surface tension[33]. While at scales longer than a mm, gravity normally dominates, surface tension typically rules when going to the micron scale. For an aquatic system like ours that is having a characteristic size (e.g. diameter) of a 30 μm, the Eötvös number suggests that surface tension is 8250 times stronger than gravity. It is therefore that water-walled microfluidics, such as presented here, might function as durable devices even at high accelerations of several Gs.

**Supplementary Movie 1: Fabrication of a water-walled microfluidic** (slowed down by a factor of 8)

**Supplementary Movie 2: Vaporization of control-group water droplets** that are sprayed on a resolution target (real time).